\documentclass[useAMS,usenatbib]{mn2e}
\usepackage[T1]{fontenc}
\usepackage{graphics,graphicx,epsfig}
\usepackage{amssymb,amsmath,amsfonts,amssymb,subfigure}
\usepackage{mathrsfs}

\title[Tidal debris from SFDM]{SFDM: A new formation mechanism of tidal debris}
\author[Victor H. Robles, L. A. Martinez-Medina and T. Matos]{Victor H. Robles\thanks{E-mail:
vrobles@fis.cinvestav.mx}, L. A. Martinez-Medina and T. Matos\\
Departamento de F\'isica,Centro de Investigaci\'on y de Estudios Avanzados del IPN, AP 14-740, 0700  D.F., M\'exico\\}

\begin{document}

\maketitle

\label{firstpage}

\begin{abstract}
Recent observations of tidal debris around galaxies have revealed that the structural properties of the spheroidal 
components of tidally disturbed galaxies are similar to those found in non-interacting early-type galaxies(ETGs), likely 
due to minor merging events that do not strongly affect the bulge region or to major mergers that happened a long time ago. 
We show that independently of merger events, tidal features like shells or rings can 
also arise if the the dark matter is an ultra light scalar field of mass $\sim 10^{-22}$eV$/c^2$.
In the scalar field dark matter (SFDM) model the small mass precludes halo formation below $\sim 10^8 M_{\odot}$ 
reducing the number of small galaxies today, it produces shallow density profiles due to the uncertainty principle in 
contrast to the steep profiles found in the standard cold dark matter (CDM) paradigm, in addition to the usual soliton solution there 
exists dark matter haloes in multistates, characterized by ripples in their density profiles, which are stable provided that
most of the halo mass resides in the ground state.  
We use the hydrodynamics code ZEUS to track the gas evolution in a background potential given by a superposition of the ground and 
first excited state of the scalar field, we study this configuration when it is initially unstable (excited state more massive 
than ground state) but by a population inversion in the states it eventually becomes stable, 
this could happen when haloes decoupled from the expansion of the universe and collapse to reach a state of equilibrium. 
We found that tidal structures like rings are formed at a particular radius as a direct consequence of the
wavelike structure of the dark matter halo, thus they possess a certain degree of symmetry that can be used to distinguish them 
from remnants formed through the usual galaxy merging mechanism.
In our simulations most of the gas concentrates in the center of the halo forming a dense and compact structure likely
enhancing star formation in this region, however, some of the gas always remains trapped between the overdensity 
regions of the background SFDM halo predicting surrounding features even in non-interacting galaxies.
In galaxies that are not strongly interacting with their neighbors, these structures may leave traces such as
radial gradients in stellar metallicity, rings of dust, or even in the form of a younger population of stars at particular radii in contrast 
to the old stellar population dominating the bulge region, our results offer a new mechanism in which features like rings and shells
can emerge purely from the quantum properties of this ultra light scalar field.

\end{abstract}


\section{Introduction}

The standard model of cosmology (or CDM) assumes that the dark matter(DM) is cold and effectively collisionless. 
It assumes a hierarchical framework for galaxy formation and throughout their evolution galaxies 
are subject to frequent collisions and interactions with nearby galaxies that leave surrounding material in the 
form of tidal debris, therefore studying tidal features can yield important clues on galaxy evolution.

Tidal features can appear in the form of complicated structures\citep{bar92}, but it is frequent to observe 
structures like tidal tails around the bulge of massive galaxies\citep{for86,ben06}, 
some early-type galaxies present shells\citep{mal80,mal83,tur99,van05,sch92,mic04} which can be seen as 
smooth arcs of ellipses centered around its host galaxy decreasing their surface brightness for larger radii,
there are also debris in the form of stellar streams\citep{mar08,mar09,mar10}.

Several analyses of the spheroidal components of tidally disturbed early-type galaxies(ETGs)\citep{tae12} suggest that structural 
properties such as the effective radius(radius at which half of the total light of the system is emitted) 
and their surface brightness are quite similar from those of the other ellipticals or the spheroids of 
galaxies with few or no signs of tidal interaction.
Other results have shown evidence of kinematically decoupled components in early-type galaxies\citep{hau06,kop00}.
Given the standard paradigm of CDM it is expected that even these ``passive'' galaxies also formed by galaxy mergers, 
then the similarity of the bulge regions for the ETGs could be due to the some minor mergers(galaxy mergers of unequal mass with mass ratio of $1/10 - 1/100$) slightly affecting the 
bulge but massive enough to still induce tidal debris, or by major mergers(mergers of comparable mass galaxies) that happened a long time ago so that now the bulge 
has relaxed.
Cosmological N-Body simulations in CDM have shown\citep{ana11,beh13} 
that merger events are frequent when galaxies are growing, however, it must be taken into account that anticipating 
the final structure around a galaxy is rather difficult given the dependence on the environment and merger history in 
the formation of each particular galaxy\citep{phi14}.

Observationally, it is not uncommon to run into difficulties identifying and studying tidal structures around galaxies, 
sometimes these structures are too faint compared to the central light distribution that a detailed analysis proves 
impossible. Nevertheless, there are observations where faint structures such as shells are clearly 
seen around galaxies at some particular radii from the center of their associated host\citep{tur99,mal83,but10,lau10,for86,sch88}.
In some situations it has been possible to identify younger stellar populations in the outskirts compared to those in the 
inner bulge\citep{kan09} probably reminiscent of a late gas accretion.
It is worth noticing that in the CDM model the spatial location of these remnants of past interaction 
is not predicted and has to be tracked with numerical simulations as it depends on the history of each galaxy.

Metallicity is another powerful tool to constrain galaxy evolution across cosmic epochs 
as it is known to be correlated to gas infall, outflows due to supernovae, and star formation rate within the galaxy. 
Recent studies of star formation focused on local group dwarf galaxies \citep{wei14a,wei14b} revealing radial 
gradients in metallicity \citep{kir13}, and most of them seem to have old stellar populations formed $\sim$ 10 Gyrs ago\citep{pie14},
followed by a more quiescent star formation likely due to loss of gas mass at the reionization epoch. 
There are high redshift galaxies ($z\sim$~3) that also show metallicity gradients, their regions of high star formation rate(SFR) 
are also of lower metallicity\citep{man09,man10,cre10} due to a large gas infall, they are sorrounded by a more enriched disc, 
in \citet{man10} the authors concluded that these observations are consistent with accretion of metal-poor gas in massive galaxies 
where a high SFR can be sustained without requiring frequent mergers, this picture was also suggested by studies of gas rich 
disks at $z\sim$ 1-2\citep{for09,tac10}.

In order to describe the debris around galaxies and stellar population it is necessary to include the gas dynamics in simulations,
Even though there is an uncertainty due to our lack of understanding of the baryonic processes, their addition is also 
required to partially reduce some discrepancies of the CDM model, for instance, the cusp-core 
problem(see \citet{blo10} for a review), the overabundance of satellites\citep{kly99,moo99,sim07,bel10,mac12,gar14} or 
the Too-Big-to-Fail\citep{boy11,gar14b} issue.
However, some of these discrepancies might also be solved assuming different properties for the 
dark matter, leading to for instance, scalar field dark matter(SFDM) \citep{sin94,lee96,guz00,mat01}, 
strongly self-interacting DM\citep{spe00,vog12,col02,roc13}, warm dark matter\citep{bod01,mac12}. 

It is of particular interest for us the SFDM alternative, here the mass of the field is assumed to be 
very small ($\sim 10^{-22}$eV/$c^2$) such that its de Broglie wavelength is of order $\sim$~kpc, relevant for galactic scales. 
The quantum behavior of the field has created much interest in the model due to its success to account for some discrepancies
mentioned above with dark matter properties only, for example, the small mass keeps the central density from 
increasing indefinitely due to the uncertainty principle in contrast to CDM simulations where supernova feedback is 
required\citep{gov10,gov12,pon12,sca13}, note that the amount of feedback required to produce a constant central density 
may be in conflict with other observations\citep{kuz11,pen12,boy14}.

An interesting property that has been noted for scalar field haloes in excited states or in a superposition of 
them is the appearance of ``ripples'' in their mass density profiles\citep{sei90,bal98,guz04,ure10,ber10,mat07,rob13}.
This property motivated us to investigate whether features like rings or shells in non-interacting galaxies 
hosted by multistate scalar field haloes could arise independently of the galaxy merger mechanism.

In the next section we give a brief overview of previous works in the scalar field dark matter model and describe 
a model for galaxy formation in SFDM haloes. 
We modify the hydrodynamics code ZEUS to implement an evolving SFDM halo in which a gaseous component 
is present, we provide the details of our simulations in Section 3. 
In Section 4 we discuss our results and section 5 is devoted to our conclusions. 

\section{SFDM}
 \subsection{Overview}

The main idea in the scalar field dark matter model\citep{sin94,ji94,lee96,guz00,mat01,hu00} considers 
a self-interacting scalar field with a very small mass, typically of $\sim 10^{-22}$eV/$c^2$, such that the quantum mechanical uncertainty principle
and the interactions prevent gravitational collapse in self-gravitating structures, thus the haloes are characterized with 
homogeneous densities (usually refered as a cores) in their centers, in general the core sizes depend on the 
values of the mass and the self-interacting parameters\citep{col86}(for a review see \cite{sua13,rin14}).

There has been several studies of the cosmological evolution that results for a scalar field mass $m\sim$10$^{-22}$eV/$c^2$, 
with this mass the cosmological density evolution of CDM can be reproduced\citep{mat01,cha11,sua11,mag12,sch14}, 
there is consistency with the acoustic peaks of the cosmic microwave background radiation\citep{rod10} and it implies a 
sharp cut-off in the mass power spectrum for halo masses below $10^{8}$M$_{\odot}$ suppressing structure formation of low mass 
dark matter haloes\citep{mar14,boz14,hu00,mat01}. Moreover, there is particular interest in finding 
equilibrium configurations of the system of equations that describe the field (Einstein-Klein-Gordon system) and of its 
weak field approximation (Schr\"{o}dinger-Poisson(SP) system), different authors have obtained solutions interpretated 
as boson stars or later as dark matter haloes showing agreement with rotation curves in galaxies and 
velocity dispersion profiles in dwarf spheroidal 
galaxies\citep{sei91,lee96,boh07,rob12,rob13,lor12,lor14,med15,die14,guz14}.
So far the large and small scales observations are well described with the small mass and thus has been taken as a prefered value 
but the precise values of the mass and self-interaction parameters are still uncertain, we believe tighter constraints can come  
from numerical simulations \citep{sch14} and comparisons with large galaxy samples which we plan to do in the future. 

Recently the idea of the scalar field has gained interest, given the uncertainty in the parameters 
the model has adopted different names in the literature depending on the regime that is under discussion, 
for instance, if the interactions are not present and the mass is $\sim 10^{-22}$eV/$c^2$ this limit 
was called fuzzy dark matter\citep{hu00} or more recently wave dark matter\citep{sch14}, 
another limit is when the SF self-interactions are described with a quartic term in the scalar field potential and 
dominate over the mass(quadratic) term, this was studied in \citep{goo00,sle12} and called repulsive dark matter or 
fluid dark matter by\citep{pee00}. 

Notice that for a scalar field mass of $\sim 10^{-22}$eV/$c^2$ the critical temperature of condensation for the 
field is T$_\mathrm{crit}\sim m^{-5/3}\sim$TeV, which is very high, if the temperature of the field is below its 
critical temperature it can form a cosmological Bose Einstein condensate, if it condenses it is called Bose-Einstein condensed(BEC) 
dark matter\citep{mat01,guz00,ber10,rob13,har11,cha11,li14}. \citet{sik09} mentioned that axions could also form Bose-Einstein 
condensates even though their mass is larger than the previous preferred value, notice that the result was contested in \citet{dav13}, 
this suggest that the condensation process should be study in more detail to confirm it can remain as BEC dark matter. 
In \cite{ure09}, it was found that complex scalar field with $m$<$10^{-14}$eV/$c^2$ that decoupled being still 
relativistic will always form a cosmological Bose-Einstein condensate described by the ground state wavefunction, 
this does not preclude the existence of bosons with higher energy, particulary in dark matter halos. 

We see that the smallness of the boson mass is its characteristic property and cosmological condensation is a likely consequence.
The preferred mass of the scalar field dark matter lies close to $\sim 10^{-22}$eV/$c^2$ satisfying the above constraint, 
although there are still uncertainties on the mass parameter, in order to avoid confusion with the known axion and help 
with the identification in future works, we find it useful and appropiate to name the scalar field dark matter candidate, 
given the above characteristics we can define it as a particle with mass $m$<$10^{-14}$eV/$c^2$, 
being commonly described by its wavefunction we choose to name this DM candidate \textit{psyon}.

It is worth emphasizing that despite the variety of names given to the model the main idea mentioned above remains the same, 
it is the quantum properties that arise due to the small mass of the boson that characterize and distinguishes this paradigm,
analoguous to the standard cosmological model represented by the CDM paradigm whose preferred dark matter 
candidates are the WIMPs(weakly interacting massive particles), one being the neutralino, 
we see that all the above regimes SFDM, Repulsive DM, Axion DM, or any other model assuming an ultra light bosonic particle
comprise a single class of paradigm, which we call \textit{Quantum Dark Matter}(QDM) paradigm. 
As pointed before, in the QDM paradigm the small mass of the dark matter boson leads to the posibility of forming cosmological 
condensates, even for axions which are non-thermally produced and have masses in $10^{-3}-10^{-6}$eV/$c^2$\citep{sik09}, 
from here we can obtain a characteristic property that distinguishes these dark matter candidates from WIMPs or neutrinos, 
namely, the existence of \textit{bosons in the condensed state}, or simply \textit{BICS}, from our above discussion the 
axion and psyon are included in the BICS.

 \subsection{Dark matter haloes of scalar field}

There has been considerable work to find numerical solutions to the non-interacting SFDM in the non-relativistic regime 
to model spherically symmetric haloes\citep{guz04,ure10,ber10,bra12,ruf69,kau68,sei91,lee10}, 
and also for the self-interacting SFDM \citep{boh07,rob12,col86,rin12,bal98,goo00}, 
it is worth noting that as mentioned in \cite{guz04} for the weak field limit of the system that determines the evolution of a spherically symmetric 
scalar field, that is, the Einstein and Klein-Gordon equations, for a complex and a real scalar field the system reduces 
to the Schr\"{o}dinger-Poisson equations\citep{arb03}.
From current bounds reported in \cite{li14} obtained by imposing that the SF behaves cosmologically as pressureless matter(dust) 
we obtained that the interacting parameter would be extremely small for the typical mass of $\sim$10$^{-22}$eV/$c^2$, 
therefore we expect that solutions to the SP system with no interactions would behave qualitatively similar to those when self-interactions 
are included, this assumption is supported by the similarity in the solutions for a small self-coupling found in other works\citep{bal98,col86,bri11},
in this work we will study the non-interacting solutions.

One characteristic feature of stationary solutions of the form $\psi(\bold{x},t)= e^{-iE_n t}\phi(r)$ 
for the SP system is the appearance of nodes in the spatial function $\phi(r)$, these nodes are associated to different energy 
states of the SF, the zero node solution corresponds to the ground state, one node to the first excited state, and so on. 
These excited states solutions fit rotation curves(RCs) of large galaxies up to the outermost measured data and can even reproduce 
the wiggles seen at large radii in high-resolution observations \citep{sin94,col86,rob13}. 
However, haloes that are purely in a single excited state seem to be unstable when the number of particles is 
not conserved(finite perturbations) and decay to the ground state with different decay rates\citep{guz04,bal98},
though they are stable when the number of particles is conserved(infinitesimal perturbations).
The ground state solution is stable under finite perturbations and infinitesimal perturbations\citep{ber10,sei90},
but has difficulties to correctly fit the rotation curves in large galaxies because its associated RC has a fast keplerian behavior 
shortly after reaching its maximum value unable to remain flat enough at large radii.

One way to overcome this problem was to consider that bosons are not in one state but instead coexist in different states
within the halo, given our intention to describe dark matter haloes we will refer to such configurations as 
multistate halos(MSHs)\citep{ure10,mat07,rob13,rob13b}.
The size of the MSH is determined by the most excited state that accurately fits the RC for large radii, 
excited states are distributed to larger radii than the ground state, and in contrast to the halo with single state 
there are MSHs that are stable under finite perturbations provided the ground state in the final halo configuration 
has enough mass to stabilize the coexisting state\citep{ure10,ber10}. 
In \cite{ber10} it was shown that MSHs can be studied in the classical approach as a collection of
classical scalar fields coupled through gravity, one field $\psi_i$ for each state, this would modify the SP system such 
that its source of energy density would be the sum of densities in each state\citep{ure10}, 
where each state satisfies its respective Schr\"{o}dinger equation and the states are coupled through the 
Newtonian gravitational potential $U$.

In spherical symmetry (in units $\hbar$=1, c=1) the SF solutions, $\Phi_n$, satisfying the Einstein-Klein-Gordon system are 
related to the SP wavefunctions through 
\begin{equation}
\sqrt{8\pi G} \Phi_n(\textbf{x},t) = e^{-imt} \psi_n (\textbf{x},t)
\end{equation}
with $m$ the mass of the SF. 
For stationary solutions we can assume wavefunctions of the form $\psi_n(\bold{x},t)= e^{-iE_n t}\phi(r)$, then the 
SP system reads
\begin{equation}
\label{sp}
 \nabla^2 \phi_n = 2(U- E_n) \phi_n \nonumber
\end{equation}
\begin{equation}
 \nabla^2 U = \sum_n |\phi_n|^2 
\end{equation}
with $\nabla^2$ the Laplacian operator in spherical coordinates, and $E_n$ the energy eigenstates.

We consider the case of a MSH with only the ground and first excited states given that
for this MSH \cite{ure10} reported the critical value for its stability under small perturbations. If the ground state 
has $N^{(1)}$ particles and there are $N^{(2)}$ in the excited state, the authors found that MSH would be stable under 
small perturbations provided the fraction 
\begin{equation}
\label{eta}
 \eta:= \frac{N^{(2)}}{N^{(1)}} \leq 1.2=\eta_{max}.
\end{equation}
In all the configurations studied in which $\eta$ is larger than the maximum for stability $\eta_{max}$, the induced instability 
causes the configuration to lose a few particles and a remarkable effect takes place, the excited state does a fast transition to 
the ground state and viceversa, that is, the populations of the states are inverted such that the final $\eta$ complies 
with the stability condition (\ref{eta}), after the inversion the MSH approaches a stable configuration where the particle 
number remains constant for each state, after the transition the particles that were initially in the excited state 
are now in the ground state where they have redistributed and become more compact than in their original distribution.

This population inversion(PI) is a characteristic feature of MSHs and as we will show it can have consequences in the gas 
distribution and on the accretion rate during the period of galaxy formation. In fact, given the cut-off in the mass power
spectrum due to the small mass of the psyon we expect a delay in the first structures to collapse compared to the those 
found in CDM\citep{mat01,mar14,boz14}, this was confirmed with cosmological simulations for a mass 
of $8\times 10^{-23}eV$\citep{sch14}, therefore, this intrinsic change of states within MSHs at high redshift 
might have observable consequences in the gas and stellar properties of the galaxies that will form in such potential wells.

\subsection{Galaxy formation scenario in SFDM haloes}

Initial fluctuations that grow due to the cosmological expansion of the universe eventually separate from it and start 
collapsing due to its own gravity, at this time (known as turnaround) the halo has a number of psyons that can be in different states, 
their values would be determined before the collapse of the configuration, and have some dependence on the its local 
environment. Depending on the number density of bosons populating the excited states we can have different fates for the haloes 
as mentioned in subsection (2.2). 

From the mentioned results of the literature and based on the successful fits to different galaxies we propose the following model of 
galaxy formation in SFDM haloes. 
The smallest and less dense systems, such as dwarf galaxies, would reside in SFDM haloes with most bosons in the ground 
state except possibly for just a few excited particles\citep{med15}, this is because the potential wells 
of these haloes, being the lowest dense systems, would be just massive enough to collapse and allow the existence of 
the state of minimum energy that would form a bound configuration. Larger configurations that had initially a larger number of 
bosons in excited states than in the ground state can undergo a population inversion and reach a stable state, 
collapse to a dense ground state or become a black hole depending on how large the fraction of particles in excited states is 
after turnaround, given the different possible outcomes for this case we expect that most galaxies are formed this way, for instance, 
ellipticals can form when haloes transition to a stable configuration to form a ground state and quickly deepen the 
gravitational potential resulting in a dense and compact structure where the bulge can form, for high surface brightness galaxies 
which usually have RCs that fall slowly after its maximum, these galaxies would likely correspond to SFDM haloes whose final stable 
configuration has a comparable number of bosons in the excited and ground states as both states would contribute evenly in the 
central region increasing the potential well where more star formation can take place, and for distances larger than a given radius 
(likely the first maximum peak in the RC) the ground state would be a small contribution to the density, thus in the outer region the baryonic 
RC would remain below its maximum. 
Another possible outcome for extended haloes is when the fraction of excited particles is below its stability threshold, 
here gravity will slowly redistribute DM bosons in the stable MSH but mostly keeping the particle number constant,
considering that psyons in the excited states are subdominant but more widely distributed than the ones in the ground state of a MSH, 
the DM gravitational potential in the center would be less dense than in the high surface brightness case so that gas accretion 
would also proceed more slowly resulting in rotation curves for the baryonic matter that slowly increase to large radii 
or with almost flat profiles, this case is quite similar to what is found in low surface brighness galaxies\citep{kuz10,kuz11,kuz11b}.

This galaxy formation scenario broadly describes the different galaxy types showing that 
only due to the dark matter properties in the context of SFDM it is possible to agree with the general features of several galaxies.
However, we do not expect this scenario to represent all galaxies, in fact, to get a better description of individual 
galaxies we require taking into account other fundamental parameters that affect their evolution, such as the environment, 
angular momentum, galaxy mergers etc., we will leave a more in dept exploration of the scenario including the full astrophysical processes for a
future work.

For now, we take the given description of elliptical galaxies to explore whether 
some substructure found in some of these galaxies can be a consequence of the properties of the background DM halo 
where such ellipticals are formed. We do not discard the possibility that shells or rings result from mergers, instead, we 
explore a new possible mechanism in which they can arise offering an alternative explanation to the
tidal debris in certain non-interacting ellipticals. We study this issue in a multistate halo.
\begin{figure} \label{fig1}
\begin{tabular}{ll}
\resizebox{110pt}{98pt}{\includegraphics{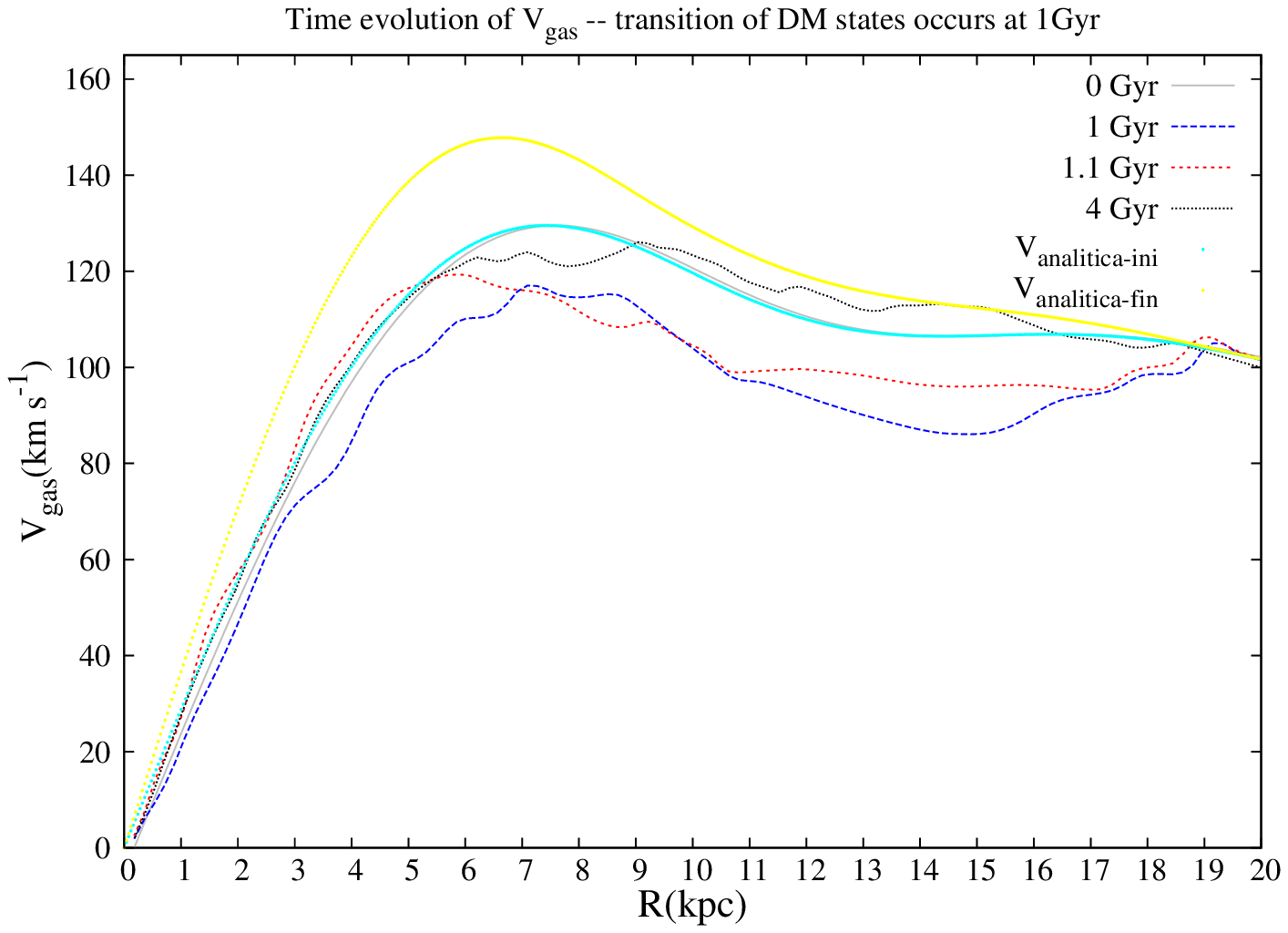}} &
\resizebox{110pt}{98pt}{\includegraphics{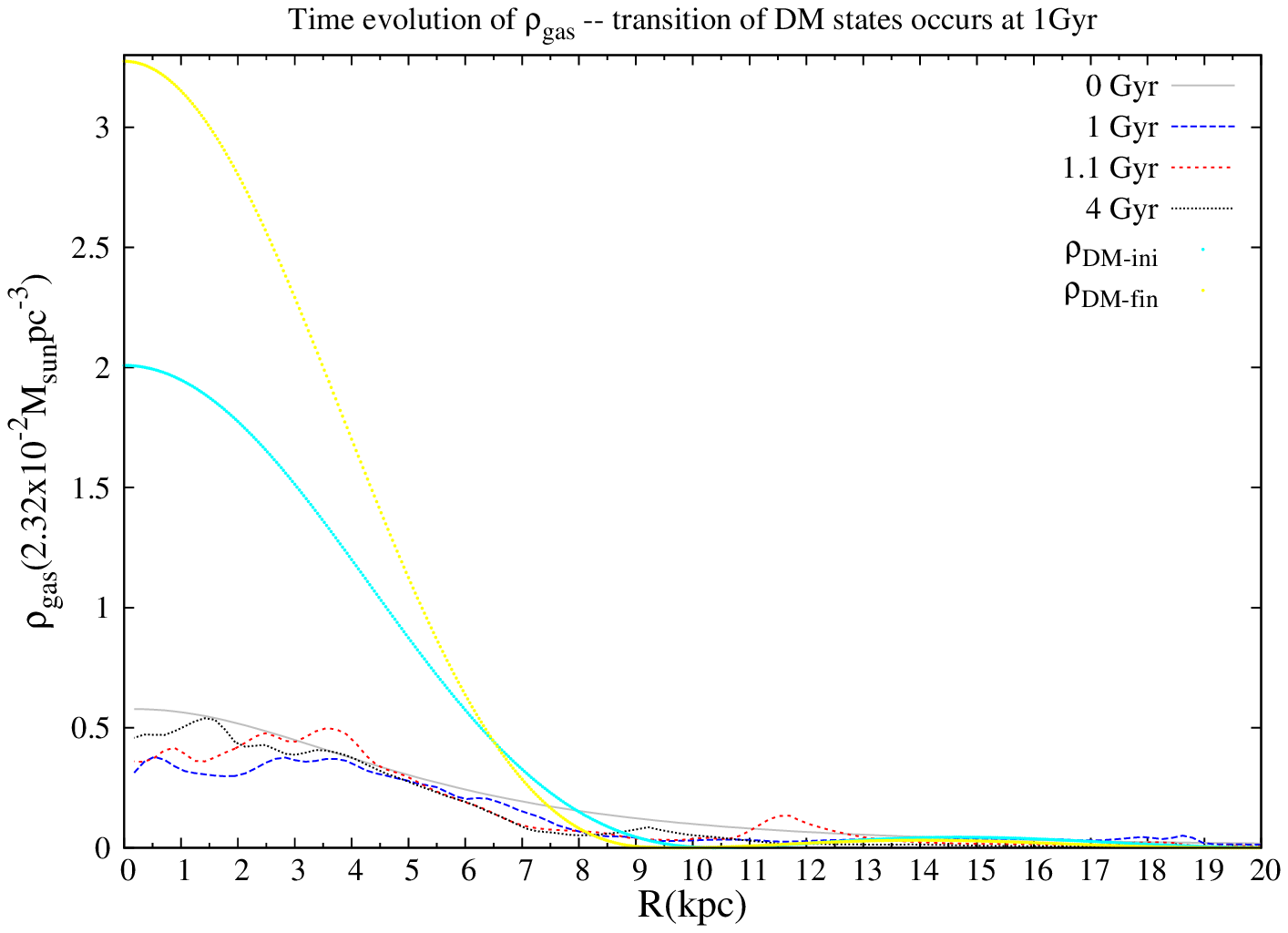}} \\
\resizebox{110pt}{98pt}{\includegraphics{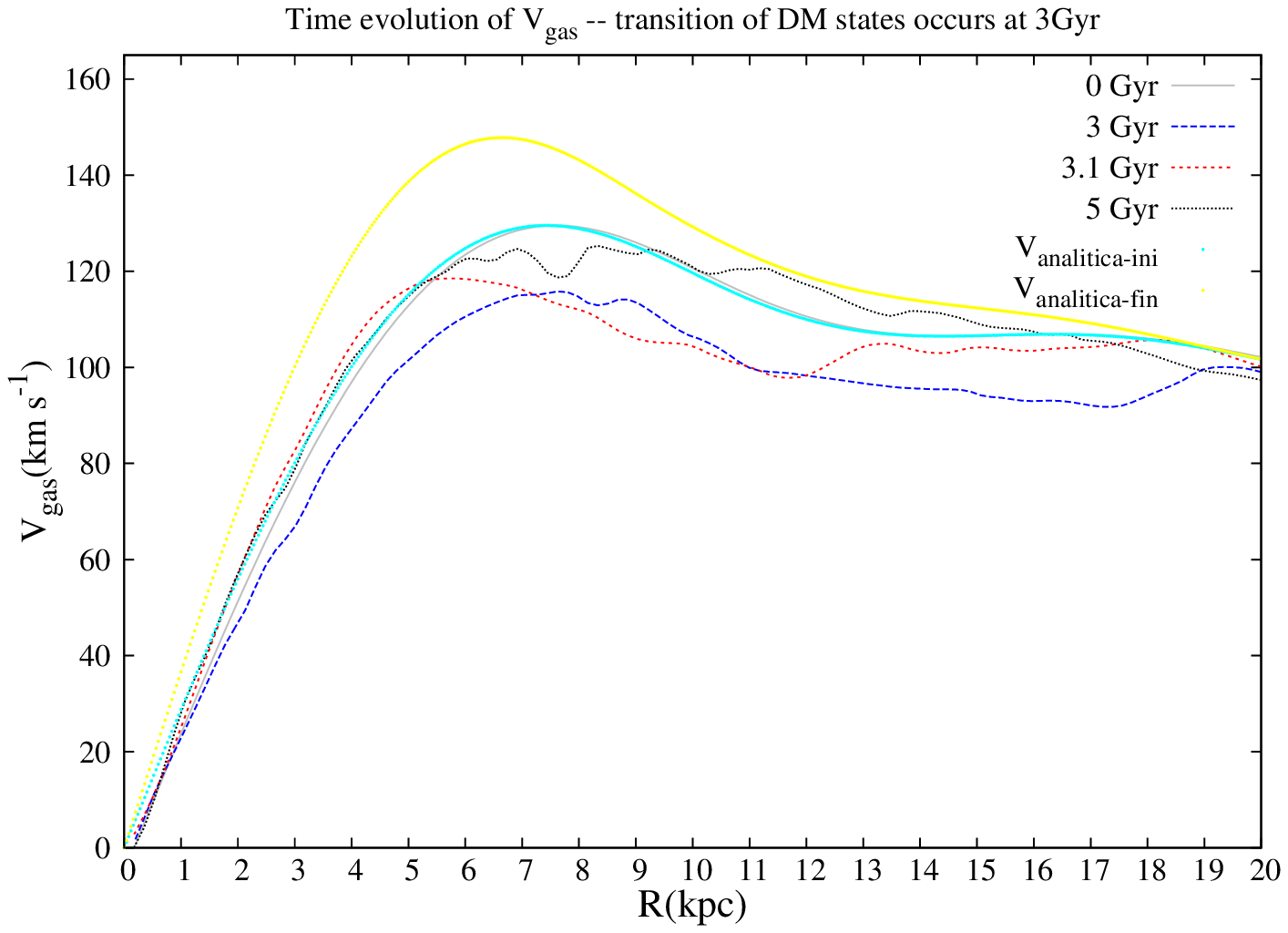}} & 
\resizebox{110pt}{98pt}{\includegraphics{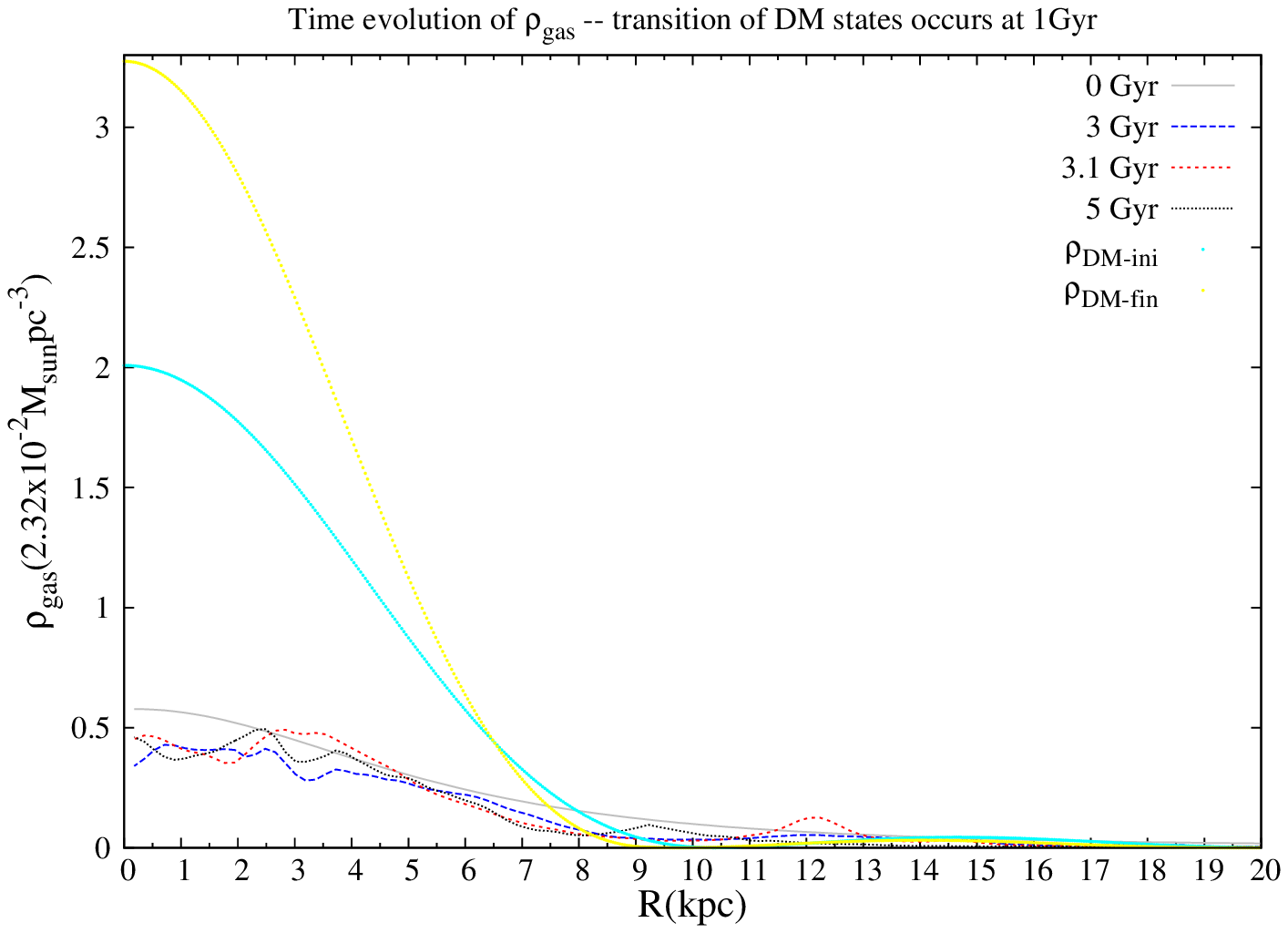}} \\
\end{tabular}
  \caption{Density profiles for dark matter and gas for simulations $A^1_1$(upper panels) and $A^3_1$(lower panels).
Shown are the initial(cyan lines) and final(yellow lines) DM profiles of the MSH composed of the ground and first states of the scalar field.
Also shown are the initial condition for the disc of gas(gray lines), its density and rotation curve just before(blue-dashed line) and 
shortly after (red-dashed line) the population inversion occurs in the SFDM halo to reach its stable configuration, the black dashed
line shows the profiles at the end of the simulations when the multistate halo is now stable.}
\end{figure}

\begin{figure} \label{fig2}
\begin{tabular}{ll}
\resizebox{110pt}{97pt}{\includegraphics{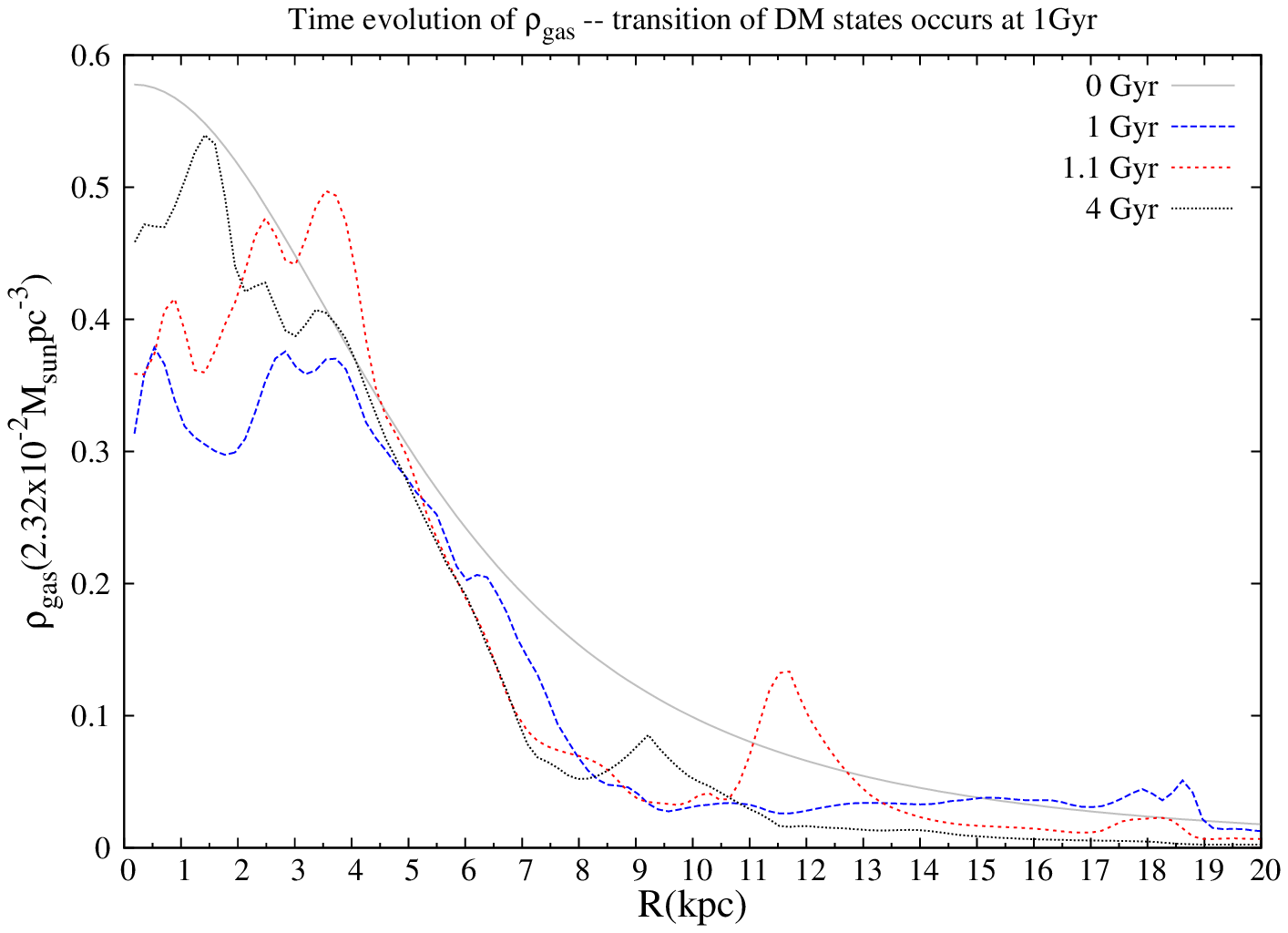}} &
\resizebox{110pt}{97pt}{\includegraphics{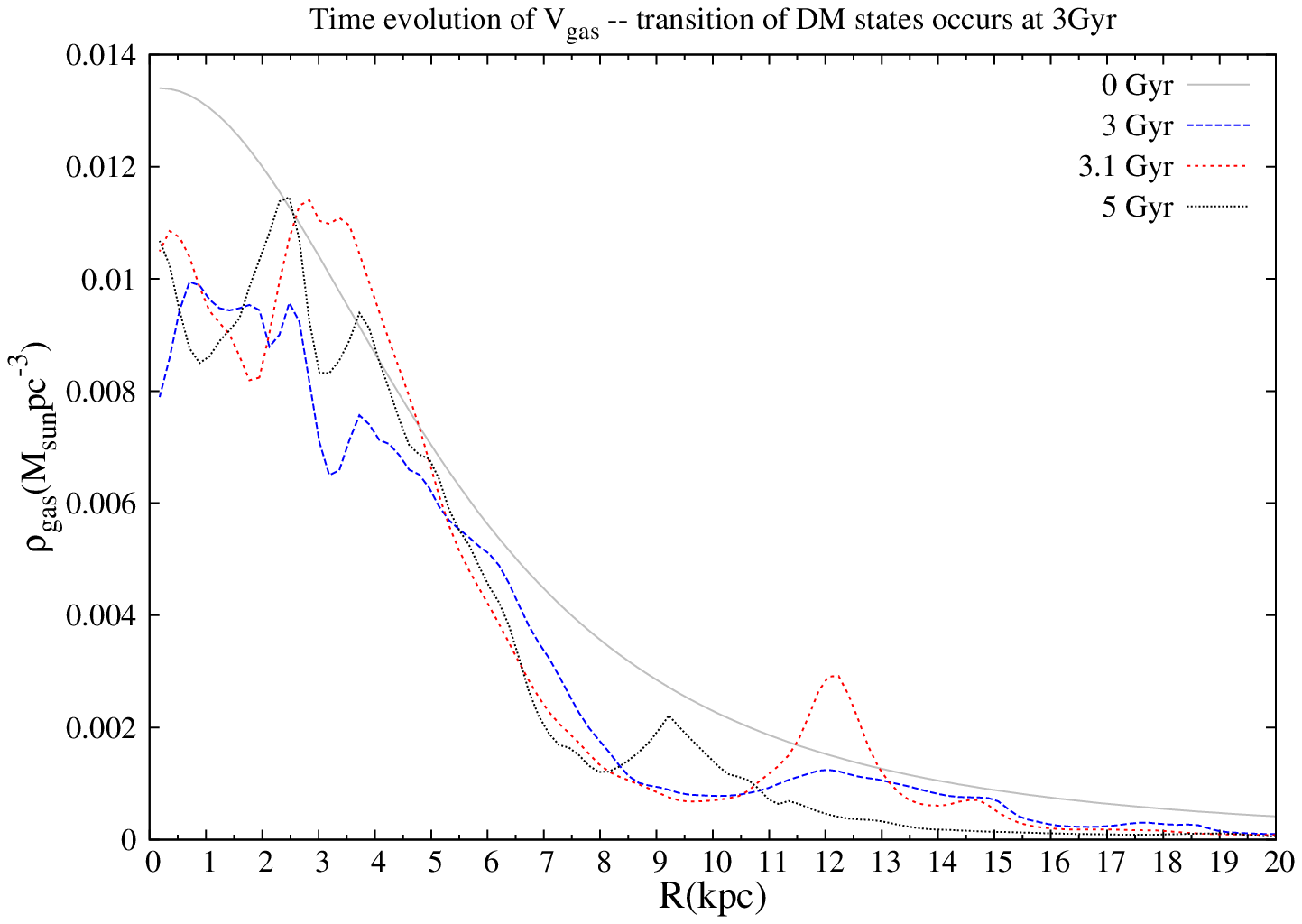}} \\
\multicolumn{2}{c}{\resizebox{165pt}{100pt}{\includegraphics{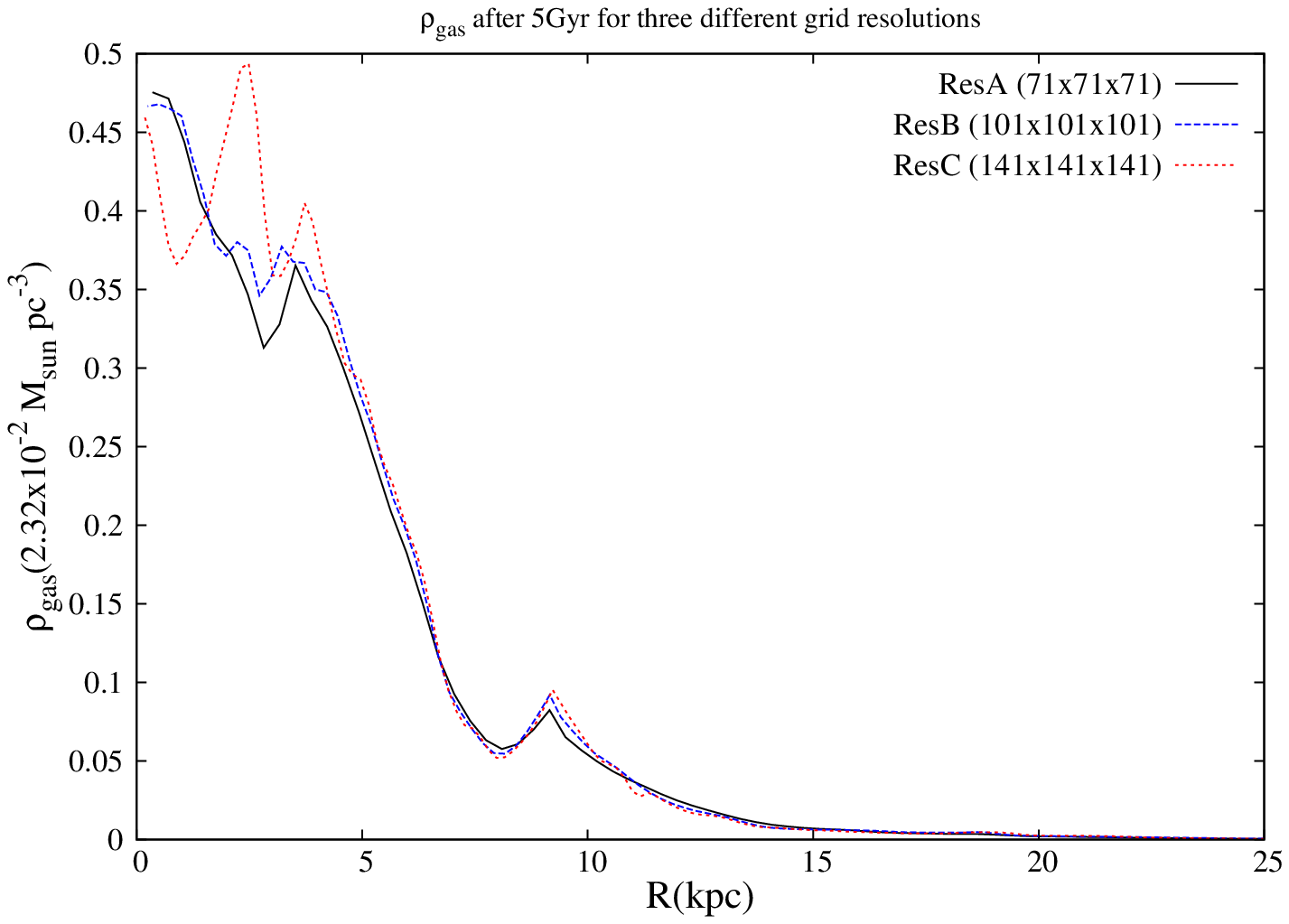}}} \\
\end{tabular}
  \caption{Density profiles for dark matter and gas for simulations $A^1_1$(upper panels) and $A^3_1$(lower panels).
Shown are the initial(cyan lines) and final(yellow lines) DM profiles of the MSH composed of the ground and first states of the scalar field.
Also shown are the initial condition for the disc of gas(gray lines), its density and rotation curve just before(blue-dashed line) and 
shortly after (red-dashed line) the population inversion occurs in the SFDM halo to reach its stable configuration, the black dashed
line shows the profiles at the end of the simulations when the multistate halo is now stable.}
\end{figure}

\section{Simulations} 

We consider a multistate halo with only the first and ground state coexisting, this is because we know the stability threshold for 
this multistate configuration. In \cite{ber10} it was seen that the larger the $\eta$ above the critical value, the faster it 
settles to its final stable configuration, to be sure that the MSH is initially in the unstable regime we pick an intermediate value 
$\eta$=$1.6$.
We explore the evolution of the gas distribution that follows a disc of gas with a Miyamoto-Nagai profile\citep{miy75}
with gas mass of $M_g= 3.9\times 10^9M_{\odot}$, and for the horizontal and vertical scale-lengths we use $a=6.5$ kpc and $b=0.5$ kpc 
respectively (we also simulated some cases with the same gas mass but using an initial spherical Plummer profile 
but found no change in the main results, there was only a slight increase in the computation time, we then report only the cases for 
an initial disc of gas).

The code we use to evolve the gaseous component embedded in the MSH is the hydrodynamics code ZEUS-MP\citep{hay06}.
It is a fixed-grid time-explicit Eulerian code, we solve for the standard hydrodynamics equation for the baryonic component 
as in \cite{med14}, but here, we modify the code to allow for the evolution of a multistate 
dark halo that undergoes a transition as described below.
In order to implent the population inversion in ZEUS we use a semi-analytical model that let us control the rate of the 
transition to a stable state helping with the interpretation of the results.
To achieve this we first use an analytical approximation to the numerical solutions obtained in the Newtonian limit 
for a confined distribution where the gravitational potential is weak\citep{ure10,guz04,rob13}, 
the scalar field density profile due to the wavefuntion in the state $j$ is 
\begin{equation}
\label{rho}
\rho_j(r)=\rho_{c,j} \frac{sin^2(j \pi r/R_j)}{(j \pi r/R_j)^2},
\end{equation}
where $j$=$1$+$n$ labels the state of the scalar field and $n$=$0,1,2,...$ is the number of nodes in the profile, 
thus $j=1$ denotes the ground state with zero nodes and $j=2$ corresponds to the first excited state, and 
the total density of the MHS would be described by $\rho(r)$=$\rho_1(r)$+$\rho_2(r)$.
In equation (\ref{rho}) $\rho_{c,j}$ represents the central density of state $j$ and $R_j$ is taken as a truncation radius of the 
corresponding state, that is, for radii grater than $R_j$ the number of particles in the state $j$ is neglected, 
thus we take $\rho_j(r)$=$0$ for all $r>R_j$ whereas for $r \leq R_j$ the density profile of the state $j$ is 
given by (\ref{rho}).

The mass profile and rotation curve for state $j$ are \citep{rob13,med14}
\begin{eqnarray}
M_j(r) &=& \frac{4 \pi  \rho_{c,j}}{k_j^2} \frac{r}{2} \biggl(1-\frac{\sin(2 k_j r)}{2 k_j r} \biggr), \label{mass}\\
V^2_j(r) &=& \frac{4 \pi G \rho_{c,j}}{2 k_j^2} \biggl(1-\frac{\sin(2 k_j r)}{2 k_j r} \biggr) \label{vel},
\end{eqnarray}
where we defined $k_j:=j \pi/R_j$. The total mass enclosed within $R_j$ is 
\begin{equation}
\label{tmass}
M_j(R_j)= \bigg(\frac{2}{\pi}\bigg) \frac{\rho_{c,j} R^3_j}{j^2}.
\end{equation}

During the population inversion in the numerical solution both states lose a few percentage of 
their initial masses(<10\% \citep{ure10,ber10}), but this loss would not substantially affect the final halo 
profilea, for this reason we will neglect the mass loss in our analytical study of the MSH. 
In order to construct the initial configuration in an unstable state we work with a mass ratio $\eta=1.6$, when the population inversion 
occurs the initial mass ratio $M^i_2(R^i_2)/M^i_1(R^i_1)$ will also be inverted and the final MSH will be stable. 
After the inversion the total initial mass of the ground state $M^\textit{i}_1(R^\textit{i}_1)$ becomes the final mass of 
the excited state $M^\textit{f}_2(R^\textit{f}_2)$ and viceversa, and the nodes in the wavefunctions will also have changed, that is 
\begin{equation} \label{invm}
M^\textit{i}_1(R^\textit{i}_1) \longrightarrow M^\textit{f}_2(R^\textit{f}_2), \hspace{2mm} 
M^\textit{i}_2(R^\textit{i}_2) \longrightarrow  M^\textit{f}_1(R^\textit{f}_1)
\end{equation}
\begin{equation}
\label{invj}
 j :1,2 \longrightarrow  2,1 = j '.
\end{equation}
We still need to specify a function to model the time evolution of the transition to the stable configuration. 
As we mentioned, for unstable haloes the transition happens really fast, during this period the central density varies until it 
reaches the new stable value, the inversion also modifies the radius of each state. In our semi-analytical approach 
the radius and density are related to the mass through eq.(\ref{tmass}),
thus we can choose to vary the radius to induce the transition, 
remembering that the transition is relatively fast, we found that a smooth function that captures the PI of the numerical evolution
and allow an interpolation between an initial and final $R_j$ is
\begin{equation}\label{rad}
R^\textit{f}_{j'}(t)= R^\textit{i}_j + \frac{\epsilon_{j',j}}{2}\bigg(1+ tanh[\alpha(t-t_{inv})] \bigg),
\end{equation}
where $\alpha$ determines the transfer rate from the initially unstable to the final stable MSH, 
$\epsilon_{j',j}=R^\textit{f}_{j'}-R^\textit{i}_j$ is a parameter that relates the initial radius to the one after the 
inversion, and $t_{inv}$ is the time where the PI is halfway to reach its final stage with $j'$ confined in $R^\textit{f}_{j'}$,
we applied a corresponding function to invert $j$ to $j'$.

Using this semi-analytical inversion model we can identify and study the dependence of the resulting 
properties of the gaseous component on the transition time, in particular the effect on the production of tidal features.
We modify the code ZEUS to implement the model and run simulations for a Miyamoto-Nagai disc of gas,
the gas has no self-gravity and is used as a tracer of the gravitational potential, the initial condition for the gas velocities 
are set directly from the background dark matter halo and therefore match the initial halo velocities(see Fig. 1).
As noted by other authors\citep{guz04,ure02,ure12} 
stationary solutions to the SP system (for a specified number of nodes) are related to each other by a scaling transformation, 
so the properties found here for our simulated parameters are expected to appear in similar configurations under a suitable scaling parameter.
Table \ref{tab1} shows the parameters of mass and radii for our MSH before and after the transition,
these are the values we will use that define the background MSH in all our runs, we explore an early and late transition 
specified by $t_{inv}$=$1$ and $3$~Gyrs respectively, and we also evolve these cases when $\alpha=1,1/4$ 
corresponding to an fast and slow inversion, we denote by $A^{t_{inv}}_{\alpha}$ a realization that had a PI with parameters 
$t_{inv}$ and $\alpha$.
\begin{table} 
  \begin{minipage}{100mm} 
  \caption{Initial and final parameters defining the MSH.}\label{tab1}
  \begin{tabular}{@{}ccccc@{}} 
  \hline
\multicolumn{3}{c}{Initial} &  \multicolumn{2}{c}{Final}\\
\hline
\multicolumn{3}{c}{$j$} &  \multicolumn{2}{c}{$j'$}  \\
\hline
     & 1& 2 & 1& 2 \\
    $M$\footnote{Mass in units $1 \times 10^{10} M_{\odot}$} & 1.844 & 2.953 & 2.953 & 1.844 \\
    $R$\footnote{Units in $kpc$} & 10 & 20 & 10 & 20 \\
\hline
\end{tabular}
\end{minipage}
\end{table}

\section{Discussion}

In Figure 1 we show the density profiles and the rotation curves for the MSH and the disc of gas comparing
the cases of a rapid early and late transitions ($A^1_1$ and $A^3_1$).
We notice the decline in the gas RC in both cases, in fact, the behavior is quite similar in both simulations, the 
gas moves according to the potential of the MSH and when the PI takes place it induces a rapid gas infall towards the center
(see red dashed line in Fig.1) seen as an bulge like region of high gas density. 
In Figure 2 we plot a zoom to the gas densities of Figure 1 where we observe a conspicuous peak that appears
as a result of the halo transition, soon after the transition a portion of the outer gas is pulled to the center and 
gets redistributed again in the now stable MSH configuration but now some of the gas remains trapped within the region of the first 
node of the MSH density profile, this results in the notable spike in the gas profiles. This density increase
appears in both early and late fast transitions showing that this feature is expected independently of when the transition to the stable 
multistate halo happened. We verified that this is not a resolution effect by simulating $A^3_1$ for three different 
grid resolutions giving the density profiles shown in the bottom panel of Fig. 2, the inner profile does change due to the higher 
resolution that captures more details of the gas while the spike at $\sim 9$ kpc is only slightly modified and at the location of 
the node in the MSH, suggesting that its origin lies in the ripple-like feature of the background MSH.
\begin{figure*} \label{fig3}
 \begin{minipage}{380pt} 
\centering
\begin{tabular}{lll}
\hspace{-1.1cm} \resizebox{155pt}{175pt}{\includegraphics{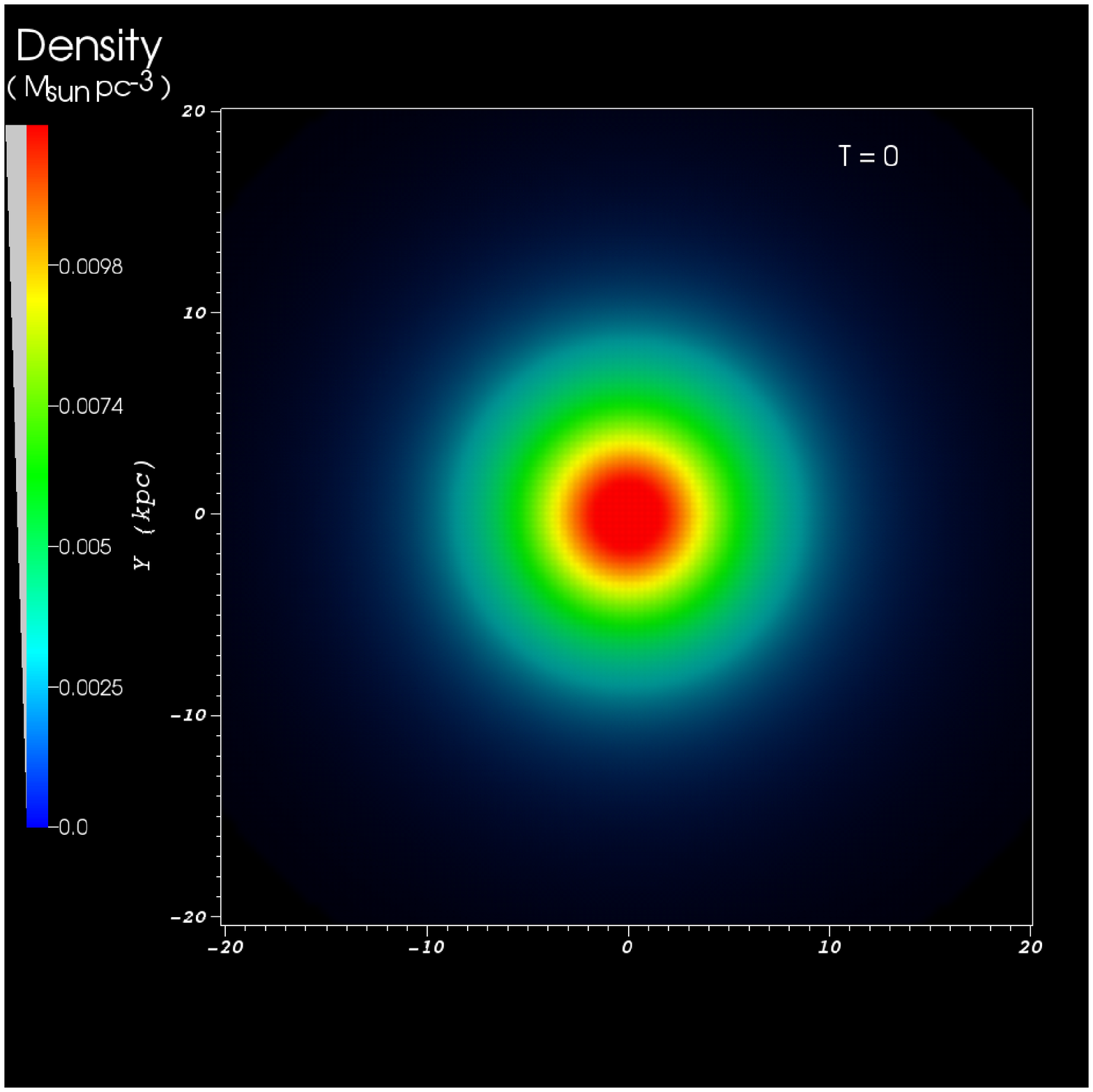}} &
\hspace{-1.1cm} \resizebox{155pt}{175pt}{\includegraphics{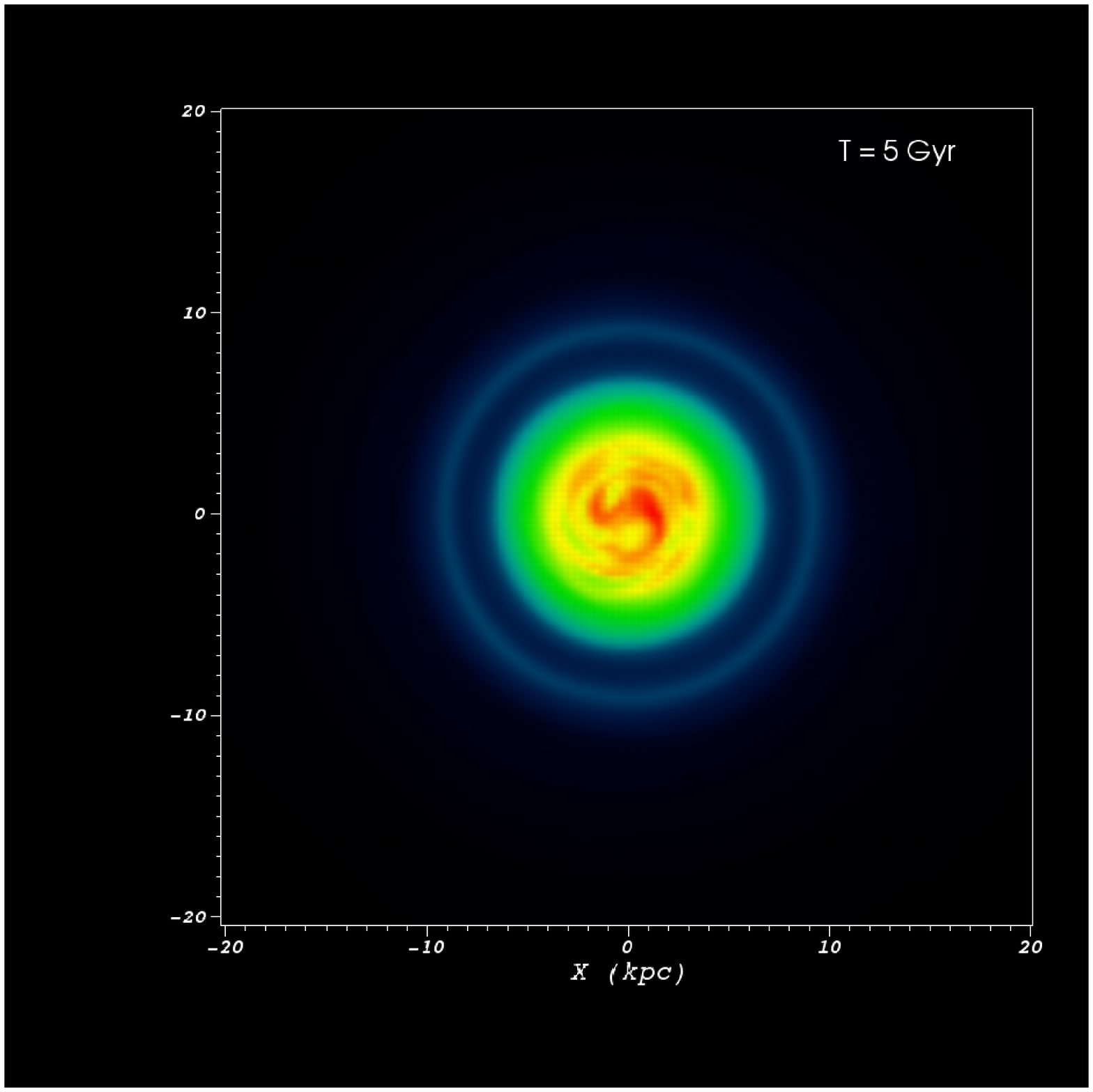}} &
\hspace{-1.1cm} \resizebox{155pt}{175pt}{\includegraphics{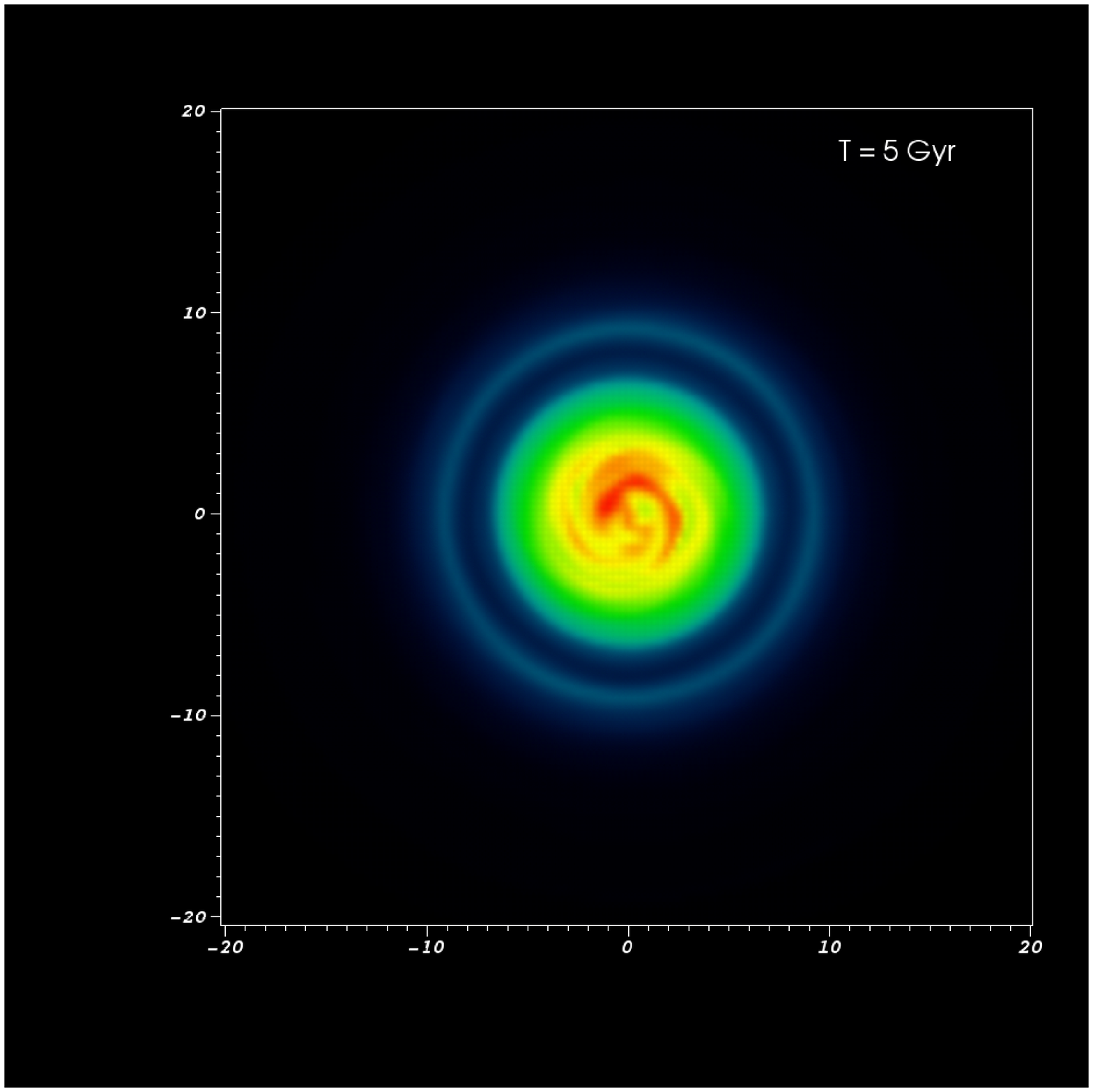}} \\
\end{tabular}
\end{minipage}
 \caption{Comparison of initial gas distribution(left) and its final stage. The central panel corresponds to a halo that had 
an early transition to its stable configuration at $t_{inv}$=1 Gyr($A^1_1$) and on the last panel(right) the transition was at a later epoch $t_{inv}$=3 Gyrs($A^3_1$).
A ring appears in both simulations at the corresponding density peak of Fig. 2.}
\end{figure*}

This surprising agglomeration of gas at $9$ kpc is nevertheless small compared to the mean central density($\approx$ 0.45M$_{\odot}pc^{-3}$) 
being $\sim 20\%$ at the peak,
the fact that it remains well after the transition is important for observations as this can be the origin of some structures around 
galaxies. In fact, in Fig. 3 we plot the gas spatial distribution at the beginning and at the end of the simulation 
with early and late halo transition, that is $A^1_1$ and $A^3_1$, and we readily observe the presence of a sorrouding ring of gas 
orbiting around the denser central concentration reminiscent of those found in certain elliptical galaxies. 
The ring keeps its location until the end of the simulations in our isolated galaxy because its origin lies 
in the shape of the background SFDM halo, moreover, the stability of the halo under small perturbations, such as accretion of small galaxies, mildy change 
the inner DM potential but may alter the ring structure in a similar way that 
the population inversion modified the initial gas profile, however, once the merging process has concluded the gas will 
start redistributing and according to our results had enough gas remain on the outskirts a ring-like structure would appear again, 
the final amplitude of the ring's density 
will depend more strongly on the details of the merging process but its location should be similar to the non perturbed halo provided the DM halo potential was 
not severely modified, as expected for minor mergers in the past or those relatively more recent, 
thus, tidal features are not exclusive of galaxy mergers when the dark matter is an ultra light scalar field.

Comparing the early and late transitions in the MSH we obtain similar structures, thus we may focus on one of them, 
we choose the case of a late population inversion($t_{inv}$=$3$ Gyrs) to explore the effect on the ring's shape under a 
fast($\alpha$=1) and slow ($\alpha$=1/4) transition. Fig. 4 depicts the comparison at the end of the simulation, 
we observe that both $A^3_1$ and $A^3_{1/4}$ form the ring as expected but the slow transition spreads the gas  
around the ring creating a broader substructure, this is consistent with our galaxy formation scenario, as the gas 
accretion proceeds slower for $A^3_{1/4}$ the ring will slowly form from the concentration of consecutively infalling gas, 
on the other hand in $A^3_1$ the rapid transition enhances the gas infall and accelerates the gathering of gas at the ring's location 
quickly depleting the rest of the gas in low density regions creating a thin and more confined ring.

Moreover, a fast gas concentration is likely to induce a period of star formation, a massive accretion in the early universe 
would generate a starburst producing several stars and winds that could drive gas away from the center, part of the latter 
will be recycled and some can be trapped in the location where the outer ring forms, there the gas can trigger star formation too 
or just accumulate as dust, the central star formation affects the metallicity in that region and if the gas infall 
stops early due to effective gas blowouts consequence of a high star formation rate the galaxy would be left with a dominant fraction of old stars. 
For a massive galaxy, part of the ejected gas cannot escape and will remain bounded scattering the inner photons and producing a luminous bulge, 
this is consistent with the formation of some ellipticals, in particular this picture could describe 
the puzzling origin of dust rings in the Sombrero galaxy(M104)\citep{baj84,ems95}, although a more careful study of this issue has to be done.

The gas fraction and nature of the dark matter halo are fundamental to determine the fate of a galaxy,
for instance, dwarf galaxies reside in SFDM haloes in the ground state and have shallow potentials, 
then if a starburst occurs early it can blow out a large portion of gas that if it is accreted much later 
the next generations of stars will differ from the old population observed as two different fractions of stellar population
where the younger one is located at larger radii due to its late accretion, analyses of stellar age 
and metallicity at different radii can be a way to distinguish between CDM predictions and the SFDM, future hydrodynamics simulations of 
SFDM will be able to confirm these predictions.

The existence of the ring is a generic feature in our simulations of isolated galaxies, notice that for an initially 
stable halo($\eta<\eta_{max}$) the ring will also appear, it suffices to observe that our final MSHs fall precisely in this regime. 
The SFDM model provides a new mechanism to form rings without major changes to the galaxy inner region mainly because it is 
where the ground state is mostly confined, in fact, the latter can exhibit a solitonic behaviour keeping its profile 
even after several mergers as shown in SFDM cosmological simulations\citep{sch14b}. 
Although perfect rings are created in our simulations, we note that this structure can be deformed by the passage of 
an accreted satellite or due to tidal forces from nearby galaxies culminating in incomplete or displaced rings, 
these tidal debris would manifest in the more common shells\citep{mal80,mal83,tur99}.

Galaxies that undergo major mergers form under violent conditions, thus the shape of their tidal structures can vary widely especially when
the interaction is still ongoing, for these latter galaxies far from equilibrium the emergence of an exotic tidal feature might 
be more accurately explained by modeling the merging galaxy and evolving it in a simulation that includes baryonic processes and a live halo. 
For the more isolated systems like the considered here, we have found that there is another mechanism that leads to outer structures 
but with the difference that they posses a certain degree of symmetry associated to the background DM halo and not to a particular merger event. 

It is interesting that the quantum properties of the psyons manifest on a macroscopic scale and have implications in observable quantities 
like gas and stars, remarkably a similar situation where quantum effects manifest on larger scales has already been observed in local experiments, 
in particular it was seen that water droplets can bounce indefinitely on the surface of a vibrating fluid bath and propel themselves 
across the surface of the fluid by virtue of their pilot-wave dynamics exhibiting quantum features like single-particle diffraction, 
tunneling, quantized orbits, and orbital level splitting\citep{wal78,cou05,cou06,pro06,edd09,edd12,for10,bus10,oza13}. 
In the SFDM context the analogue fluid would be the dark matter and the baryons would be the droplets guided by the DM, 
whether this analogy is valid deserves further consideration and will be treated in other work.

\begin{figure} \label{fig4}
\begin{tabular}{ll}
\hspace{-.3cm} \vspace{-1cm} \resizebox{140pt}{170pt}{\includegraphics{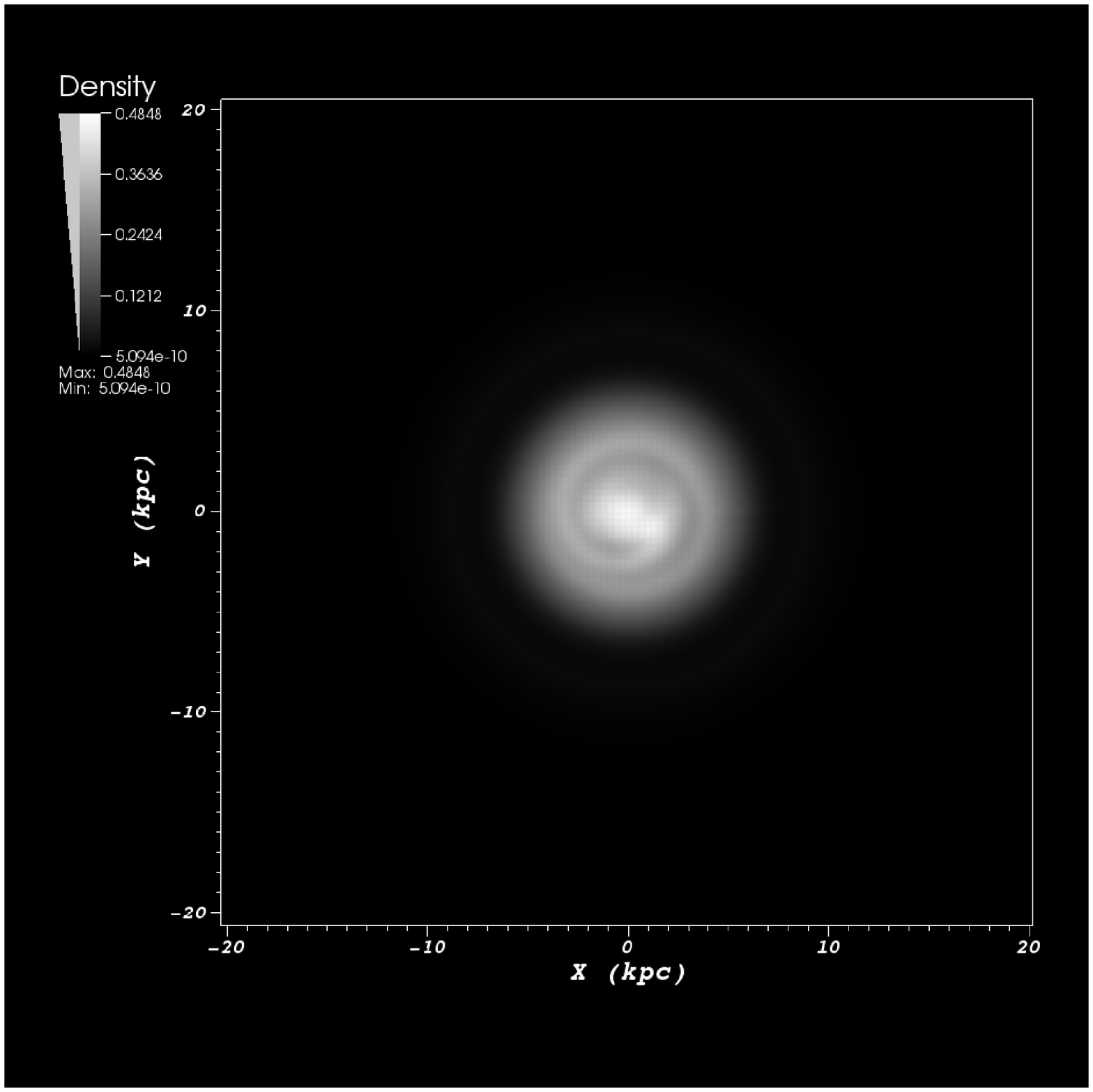}} &
\vspace{-1cm} \hspace{-1.1cm} \resizebox{140pt}{170pt}{\includegraphics{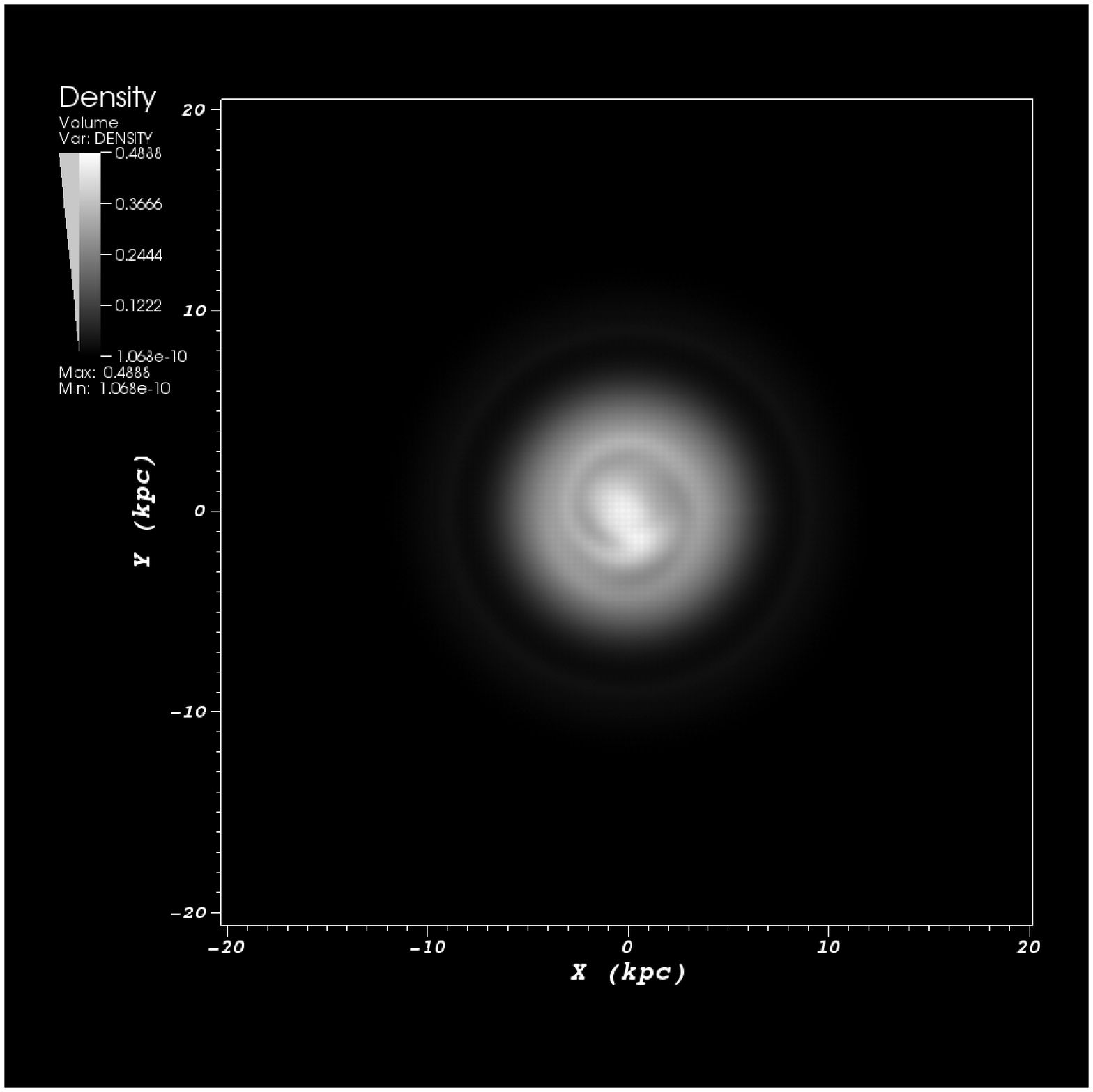}} \\
\hspace{-.3cm} \resizebox{140pt}{160pt}{\includegraphics{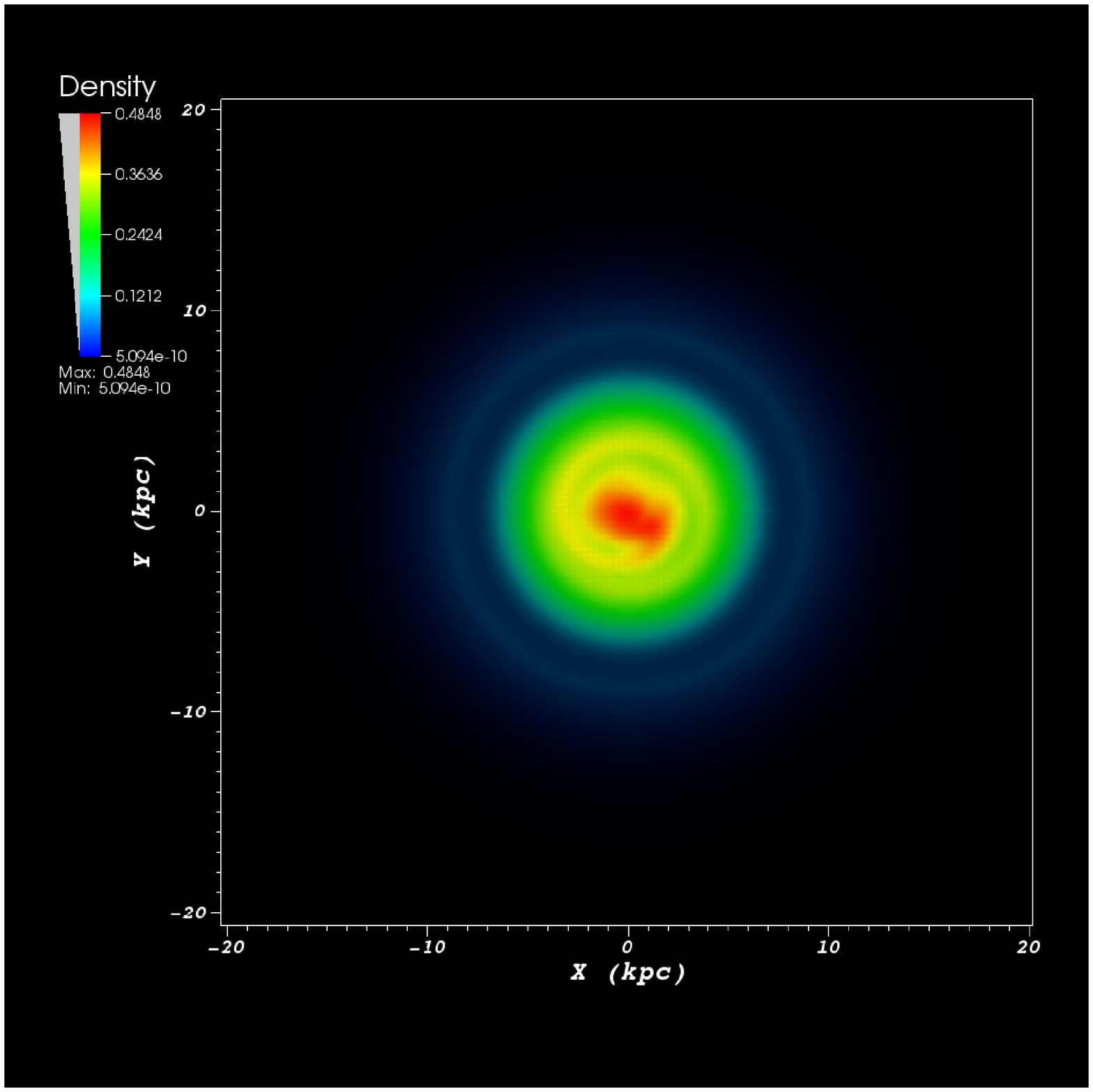}} &
\hspace{-1.1cm} \resizebox{140pt}{160pt}{\includegraphics{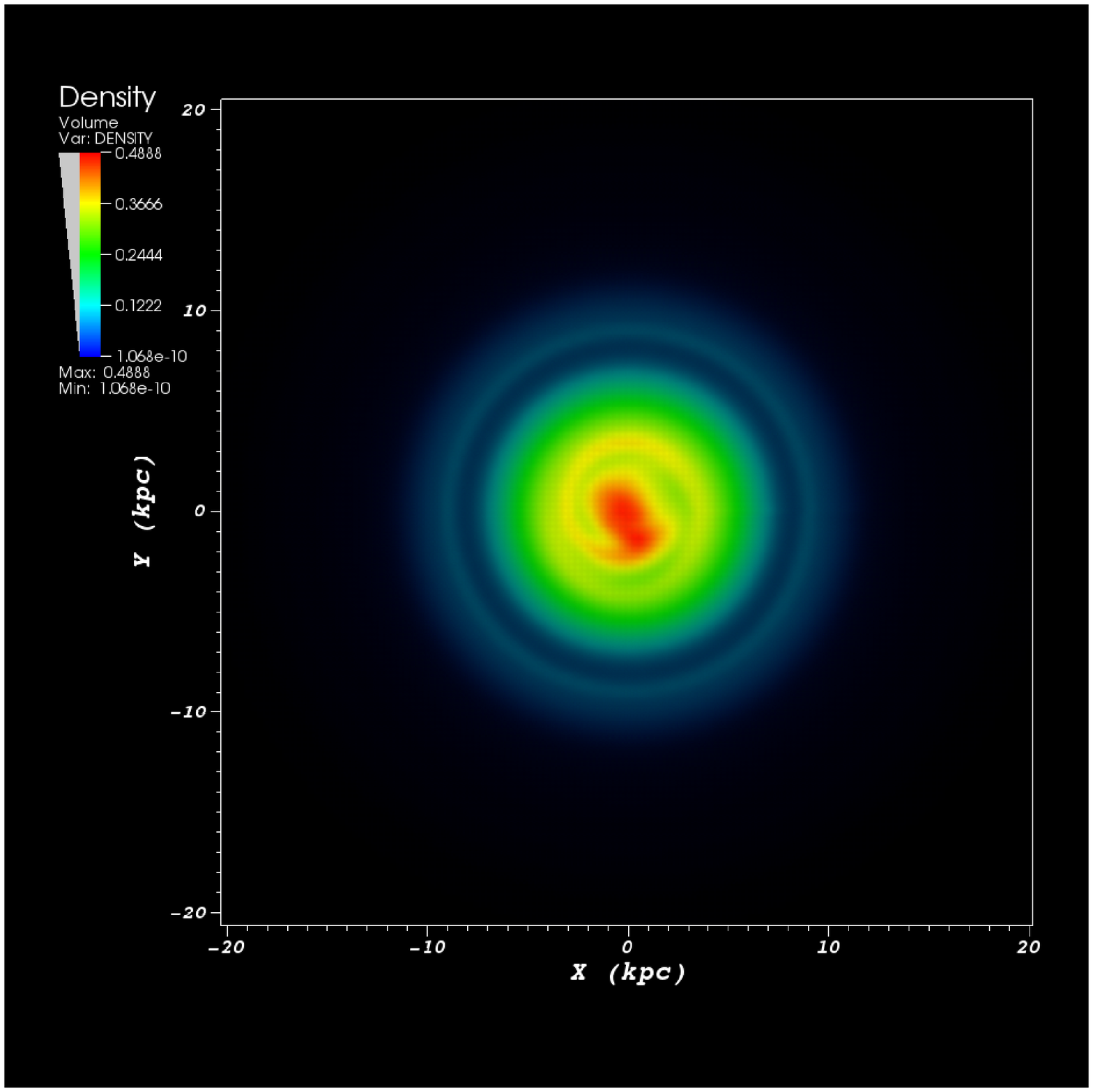}} \\
\end{tabular}
  \caption{Gas distribution after 5 Gyrs for simulations $A^3_1$(upper-left and its colored scale in the bottom-left panel) 
and $A^3_{1/4}$(upper-right and its colored version in the bottom-right panel). The ring on the left panels is more defined 
as the MSH undergoes a rapid transition to stability while for a slow transition (right panels) more particles are spread around the ring.}
\end{figure}

%

\section{Conclusions}

In this work we have seen that if the dark matter is a ultra light scalar field the internal structure of 
galaxies can reflect fossil records of their formation, in our case we study the formation of tidal structures 
similar to rings and shells in non-interacting galaxies. We discussed the nature of these structures in the context of SFDM 
and conclude that independently of galaxy mergers they can arise as a consequence of the quantum nature of the scalar field dark matter 
boson(hereafter \textit{psyons}) that form the halo where galaxies are embedded in, mainly, due to the intrinsic ripples in the 
dark matter halo density profile.

The possibility of describing tidal substructure in scalar field dark matter halos was also suggested in \cite{bra12} but under
different halo configurations and without including the evolution of a gaseous component, here we make use of spherically symmetric 
haloes in multistates because their stability has been studied and confirmed in previous works\citep{ure10,ber10}.
In addition, they possess desirable properties to describe a variety of galaxies. 
Inspired by the numerical solutions in \cite{ure10}, where they reported the maximum fraction of psyons in excited states 
(eq. \ref{eta}) to form a stable multistate halo composed of the ground state and first excited state, 
we propose a semi-analytical model to describe a similar halo configuration and study the 
distribution of a gaseous disc that is contained within such scalar field DM halo. 
The halo is set to be initially unstable by choosing a fraction of excited boson above the stability limit, this 
induces a transition to reach a final stable configuration affecting the initial gas distribution, 
after the transition the population of the states is inverted and the halo stabilizes once the 
ground state dominates the halo mass that by gravity will be confined later in the center of the halo. 
At the end of the simulations ($5$ Gyrs) part of the gas keeps orbiting in the form of a ring outside the main bulge-like region,
the structure of the ring is not significantly different when the halo transition happens early ($t_{inv}$=$1$ Gyrs) or 
late($t_{inv}$=$3$ Gyrs) but the ring broadens if the transition is slow and becomes more defined for fast transitions. 
Due to the lack of stability parameter for MSH in higher energy states we explore the first non trivial superposition of multistates, 
given the positive results it seems that tidal structures would also be found at larger radii for SFDM haloes that have boson 
in higher excited states, although they would become fainter with distance, this would agree with shells that appear to be distribited 
at larger radii. 

The tidal features in this work have a distinct origin from the usual tidal stripping of galaxies, they emerge from the quantum properties
of the psyons that compose the dark halo, the presence of the symmetric ring is due to the spherically symmetry hypothesis,  
deviations from this symmetry could result in incomplete rings and shells although a certain degree of symmetry should remain 
which distinguishes to remnants that come from recent galaxy mergers that depend on the path of the accreted satellites and 
does not necessarily leave tidal debris of a particular symmetry, our results suggest that outer regions of dark haloes may 
not be totally smooth. Exploring in more detail the regions where shells and rings form could also help to put constraints 
on their origin, their formation, and serve to test galaxy and halo formation scenarios such as the quantum dark matter(QDM) paradigm.

Additionally, we have provided a classification for the dark matter models that are based on the assumption that the dark matter 
is a scalar field of a very small mass that proves useful for future references and that helps to establish the main feature 
of the particle candidates of this paradigm. We formulate a galaxy formation scenario based on known results of halo formation in the SFDM context 
that can explain qualitatively the general properties of the different galaxy types in the universe and applied to interpret 
our findings.
Given the success of the SFDM model to account for the apparent discrepancies of the standard CDM model with only dark matter 
properties, it will be desirable to include astrophysical processes in simulations to get a fair comparison with state-of-the-art 
CDM simulations. The QDM paradigm is indeed an interesting possibility that deserves further study.

\section*{Acknowledgments}
This work was partially supported by CONACyT
M\'exico under grants CB-2009-01, no. 132400, CB-2011, no. 166212,  and I0101/131/07 C-234/07 of the 
Instituto Avanzado de Cosmologia (IAC) collaboration (http://www.iac.edu.mx/). 
Victor H. Robles and L. Medina are supported by a CONACYT scholarship.

\end{document}